# Automatic Streetlights that Glow on Detecting Night and Object using Arduino

Zain Mumtaz[1], Saleem Ullah[1], Zeeshan Ilyas[1], Shuo Liu[2], Naila Aslam[1], Jehangir Arshad Meo[3], Hamza Ahmad Madni[2, 4*]

[1]Department of Computer Science, Khwaja Fareed University of Engineering & Information Technology, Rahim Yar Khan, 64200, Pakistan
[2]State Key Laboratory of Millimeter Waves, Department of Radio Engineering Southeast University, Nanjing 210096, China
[3]Department of Electrical Engineering, COMSATS University Islamabad, Pakistan
[4]Department of Computer Engineering, Khwaja Fareed University of Engineering & Information Technology, Rahim Yar Khan, 64200, Pakistan

**ABSTRACT** Our manuscript aims to develop a system which will lead to energy conservation and by doing so, we would be able to lighten few more homes. The proposed work is accomplished by using Arduino microcontroller and sensors that will control the electricity based on night and object's detection. Meanwhile, a counter is set that will count the number of objects passed through the road. The beauty of the proposed work is that the wastage of unused electricity can be reduced, lifetime of the streetlights gets enhance because the lights do not stay ON during the whole night, and also helps to increase safety measurements. We are confident that the proposed idea will be beneficial in the future applications of microcontrollers and sensors etc.

**INDEXED TERMS** Automation, Switching, Energy conservation, Arduino, Sensors.

Corresponding author: H. A. Madni (email:101101770@seu.edu.cn).

# I. INTRODUCTION

Automation systems [1] are being preferred over the manual mode because it reduces the use of energy to saves energy. These automation systems play an essential role in making our daily life more comfortable and facilitate users from ceiling fans to washing machines and in other applications [2]. Among all exciting applications, street lights play a vital role in our environment and also plays a critical role in providing light for safety during night-time travel. In this scenario, when the street lights are in working functionality over the whole night that consumes a lot of energy and reduces the lifetime of the electrical equipment such as electric bulb etc. Especially in cities' streetlights, it is a severe power consuming factor and also the most significant energy expenses for a city. In this regard, an intelligent lighting control system can decrease street lighting costs up to 70% [3] and increase the durability of the equipment.

The traditional lighting system has been limited to two options ON and OFF only, and it is not efficient because this kind of operations meant power loss due to continuing working on maximum voltage. Hence, wastage of power from street lights is one of the noticeable power loss, but with the use of automation, it leads to many new methods of energy and money saving. In this regard, controlling lighting system using Light Dependent Resistor (LDR) [4], IR obstacle detector sensor [5] and Arduino [6] together is proposed in the past [7-10]. In the meanwhile, the importance of smart light system has motivated a lot of studies and the series of research work has been done [7-20]. In previous works, the street light systems are based on LDR [8-13], and most of them are passive infrared receiver based systems that are controlled with timers and analog circuits. Sun tracking sensors [21] are also utilized to power OFF the street lights by the detection of the sunlight luminance. Furthermore, street light control with the use of solar energy [11], and ZigBee based system to control street light [22] have also been implemented. Distinguished from turning ON/OFF the electricity, another approach is introduced to dim the light [10] in fewer traffic hours that might be useful to reduce the power consumption, but the electric bulbs are in continuous usage condition. To the best of our knowledge, a need is still existed to design a system that controls the dim light, connect the power ON/OFF with the vehicle's motion detection, calculate the total number of vehicles passed through the road, and control the entrance gate at night to reduce criminal activities.

The most natural solution is to control the street lights according to the outside lighting condition. This is what our paper is aiming for in smart lighting system in which the street lights will be turned OFF when there are no motion detections or day-time, otherwise the lights will be remained Dim/ON. Our proposed design is aimed at efficiently replacing any light systems that are manually controlled, and this is accomplished with the properly arrangements of microcontroller Arduino Uno, IR obstacle avoidance sensor, LDR, and Resistors. In this scenario, when the intensity of sunlight impinges with LDR, street lights can be further controlled as per the desired requirement, automatically. Most importantly, a counter is set to count the number of vehicles/objects passing through the road, which will be displayed on the serial monitor of Arduino IDE [6]. Moreover, the high-intensity discharge street bulbs [23] are replaced with LEDs to further reduce the power consumption. An automatic street light system does not help us in reducing the power consumption only, but also to reduce accidents, criminal activities and maintenance costs.

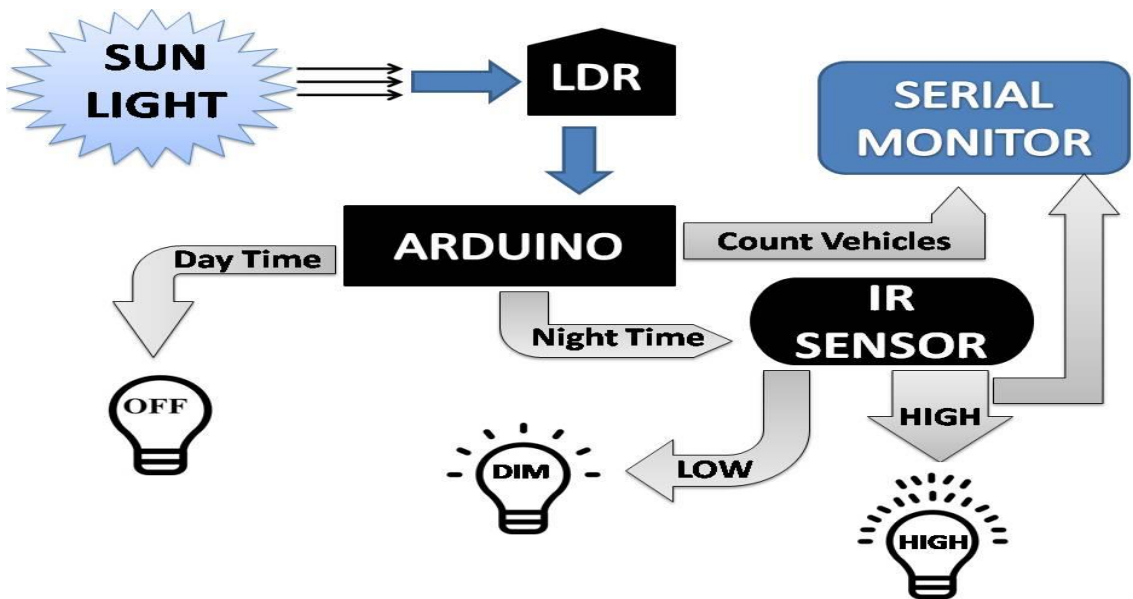

**Figure 1.** The architecture design of automatic street light control system.

For the simplicity of discussion, Fig. 1 illustrates the overall working mechanism and the features of the proposed lighting concept. Firstly, LDR will sense the intensity value of sunlight and send it to Arduino. Arduino will judge if the received value is above the threshold level (which is set independently by the user from the discrete value: 0-2023), then it will consider it as day-time and LEDs will remain OFF, or if the received value below the threshold level, Arduino will consider it as a night-time. In the night-time, if the value of IR obstacle detector sensor is LOW and detects no object, then DIM LEDs (half of its maximum voltage) will glow, or if IR obstacle detector value is HIGH and detects any object, then HIGH LEDs (full of its maximum voltage) will glow. Arduino will also count the total number of vehicles that crossed the street in the night-time with the help of IR obstacle detection sensor and will demonstrate it to the serial monitor.

Multiple electronic components are used for building electronic circuits. Our proposed circuit designs contain these components that are described below in table 1:

| Components | Specifications |
|---|---|
| 1. LDR [3] | Voltage: DC 3-5V, 5mm,1.8 gm. |
| 2. Arduino Uno [5] | 22 pins, operating voltage 6-20V |
| 3. LEDs [6] | 5 mm , operating voltage 5V |
| 4. IR obstacle avoidance sensor [4] | Voltage: DC 3-5V, Range 2-30cm, Angle 35 |
| 5. Resistors [25] | 100 ohm, 220 ohm |

**Table 1.** Specification of electronic components used in to design the proposed system.

**A. Light Dependent Resistor (LDR)**

LDR is a Light Dependent Resistor (Fig. 2a) whose resistance is dependent on the light impinging on it. The resistance offered by the sensor decreases with the increase in light strength and increases with the decrease in light strength. This device is used for detection of day-time and night-time because when sunlight falls on it, it will consider as day-time, and when there is no sunlight falls on it, it will be regarded as a night, as shown in Fig. 2b. These are very beneficial, especially in light/dark sensor circuits and help in automatically switching ON /OFF the street lights.

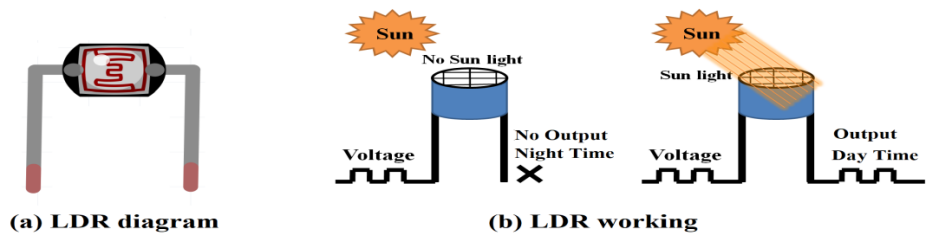

(a) LDR diagram    (b) LDR working

**Figure 2.** LDR symbol and its working phenomenon [4].

**B. Arduino Uno**

As shown in Fig. 3, the Arduino Uno [6] is a microcontroller board which is based on the ATmega328 series controllers and has an IDE (Integrated Development Environment) for writing, compiling and uploading codes to the microcontroller. It has 14 digital input and output pins (of which 6 are PWM) and 6 analogue inputs for communication with the electronic components such as sensors, switches, motors and so on. It also has 16 MHz ceramic resonators, a USB connection jack, an external power supply jack, an ICSP (in-circuit serial programmer) header, and a reset button. Its operating voltage is 5v, input voltage 7 to 12v (limit up to 20v).

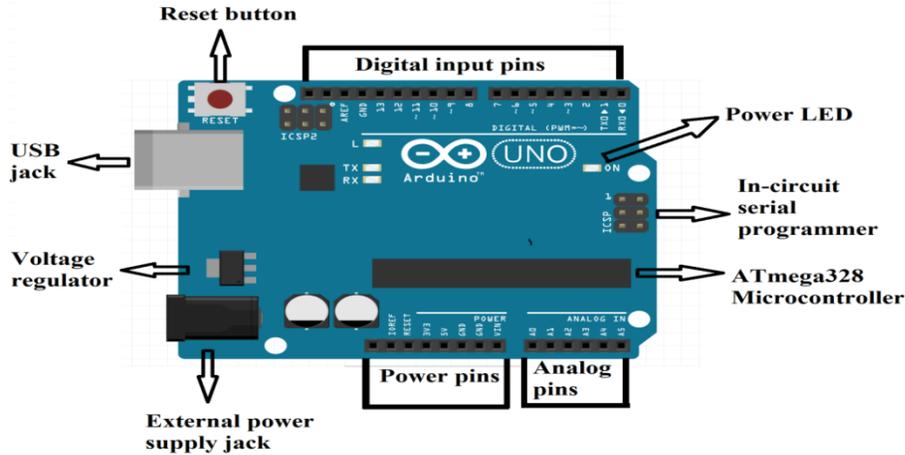

**Figure 3.** Arduino Uno board description [6].

### C. LEDs

A LED (light-emitting diode) is a PN junction diode which is used for emitting visible light when it is activated, as presented in Fig. 4. When the voltage is applied over its elements, electrons regroup with holes within the LED, releasing energy in the form of photons which gives the visible light. LEDs may have the Dim/full capability.

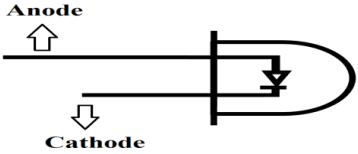

**Figure 4.** LED circuit diagram [7].

### D. IR Obstacle Avoidance Sensor

An obstacle avoidance sensor consists of an infrared-transmitter, an infrared-receiver and a potentiometer for adjusting the distance, shown in Fig. 5a. Whenever an object passes in front of a sensor, the emitted rays hit the surface of an object and reflect to the receiver of the sensor so it will consider this as a motion (as shown in Fig. 5b). It is a heat sensitive sensor and used for detection of motion.

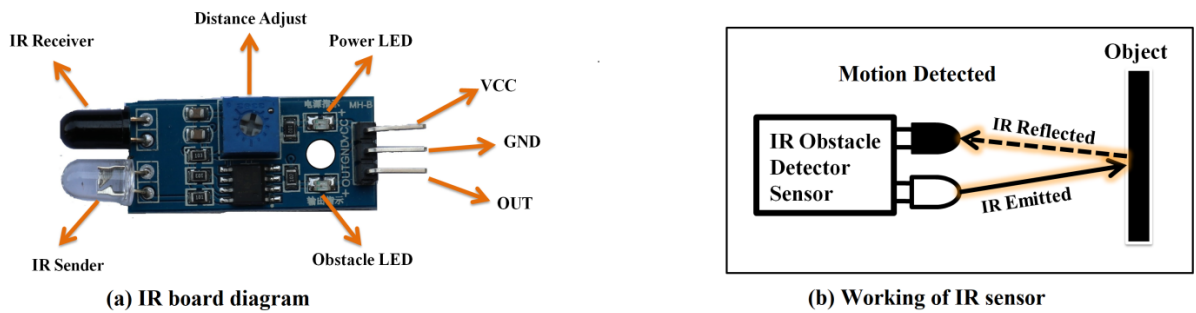

(a) IR board diagram  (b) Working of IR sensor

Figure 5. IR obstacle detector sensor diagram and working [8].

E. Resistors

A resistor is a passive electronic component, used with other electronic components such as LEDs and sensors to prevent or limit the flow of electrons through them as illustrated in Fig. 6. It works on the principle of Ohm's law which prevent overflow of voltage.

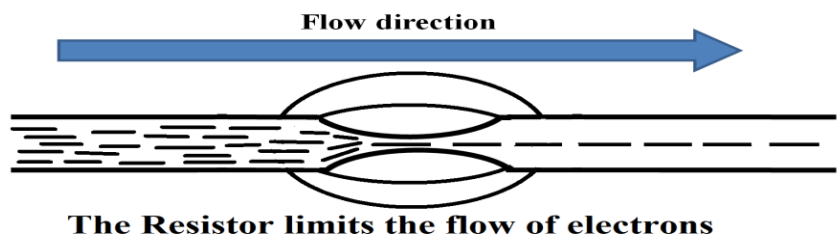

Figure 6. Working principle of resistor [26].

II. DESIGNING METHODOLOGY

A. Object Independent Automation System

Fig. 7 shows the circuit design of automatic street light control system based on vehicle detection using Arduino Uno having feature of Dim light capability. In this task, 01 LDR sensor, 12 LEDs, 13 resistors, 03 IR obstacle detector sensors and 01 Arduino UNO have been used.

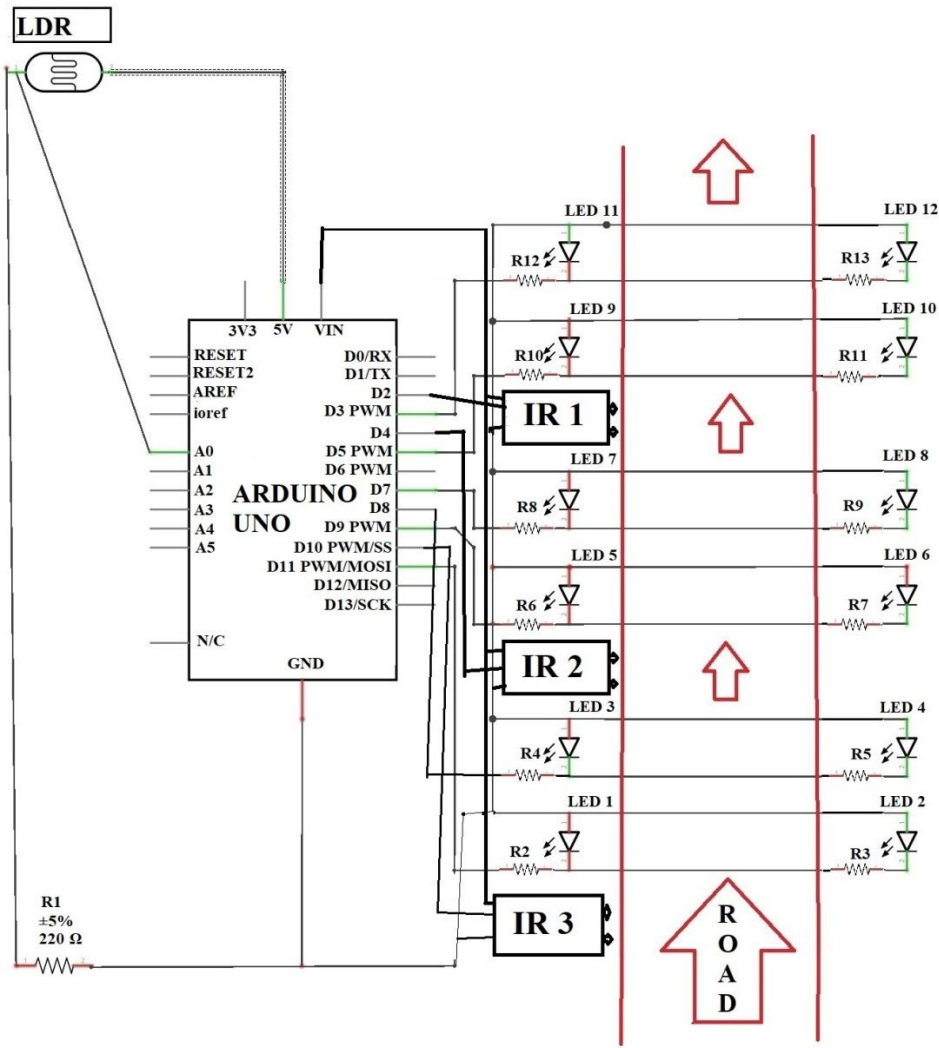

**Figure 7.** Circuit design of automatic street light control system with the Dim light capability.

One leg of LDR sensor is connected to Arduino analog pin number A0 and another leg to VCC pin and same with a resistor to the ground port of Arduino. In addition, the threshold value is adjusted to 10 from the discrete values (0-1023) for understanding whether it is day or night. After that, all the positive terminals of the LEDs are connected with resistors to pin number 3, 5, 7, 8, 9 and 11, depicting the streetlights as the outputs of the Arduino signals. Furthermore, connected the ground of all the LED's to Ground port as per the circuit diagram shown in Fig. 7. The IR obstacle avoidance sensors are connected to the Arduino port from pin number 2, 4 and 10, respectively, which is the input signal to the Arduino board. Similarly, the ground of all the IR obstacle avoidance sensors are connected to GND port and all VCC of IR obstacle avoidance sensors are attached to Arduino 5V pin. Initially, set the IR obstacle avoidance sensors to HIGH at the start if there is no motion.

After connecting all these devices to the corresponding pins in Arduino according to Fig. 7, the Arduino Software from the official website "www.arduino.cc" is downloaded and installed. Then Arduino Uno is connected to the computer using the USB cable and installed the driver software on the computer to write, compile and run the software code on Arduino software.

**B. Results & Discussions**

In the beginning, the LDR sensor will sense the light intensity in the atmosphere at that time and consequently sends the data to Arduino. After receiving the data, Arduino will convert it into different discrete values from 0 to 1023 (In which 0 represents maximum darkness and 1023 represents maximum brightness) and then it will adjust the output voltage accordingly from 0 to 2.5v/5v (Dim/High) depending upon the received value (0-2023) by comparing with threshold value. So, the output will be 2.5v in the complete darkness (night time) if the received value is less than the threshold value. As a result, Dim LEDs will glow that is the half of maximum brightness, and when there is completely shine (daytime), the received value will be higher than the threshold value, and the output voltage would be 0v resulting the LEDs to be entirely switched OFF.

Initially, the IR obstacle detection sensor will be HIGH. So, when there is no vehicle/obstacle in-front of the sensor, IR Transmitter does continuously transmit the IR light. Whenever, a car or any other object blocks any of the IR sensors, then the emitted rays will reflect the IR receiver after hitting the object, then microcontroller will sense it as a motion. In simple words, when any object passed in front of the first IR sensor, the corresponding LEDs will be turned from DIM to HIGH (5v) by the microcontroller. As the object moves forward and blocks the next IR sensor, the next three LEDs will be turned to HIGH from DIM, and the LEDs from the previous set is switched to DIM from HIGH. The process continues this way for the entire IR obstacle detector sensors and LEDs. These kinds of application can be implemented in the headlights of vehicles, street lights, parking lights of hotels, malls and homes, and it can be very beneficial.

Fig. 8 shows the result diagrams of automatic streetlights that turn to DIM at night and HIGH on vehicle movement using Arduino Uno. Fig. 8a represents the daytime with no LEDs are glowing after measuring the sensed intensity value of sunlight with the threshold value by the LDR sensor. In the meanwhile, Fig. 8b shows the nighttime because the sensed intensity value of sunlight by LDR is below than the threshold value (10) and there is no motion detected by any of IR sensors, so as a result, the DIM LEDs are glowing. Moreover, the beauty of the proposed model can be seen in Fig. 8(c-d) with the motive that only those LEDs will glow higher whose will detect the object's presence and the remaining LEDs will keep maintain their DIM state. As an example, in Fig. 8c, the first set of LEDs are glowing HIGH and remaining are in DIM mode because the sensed intensity value of sunlight by LDR is below then the threshold value so, it considered nighttime and, there is an object detected by the first IR sensor. Moreover, when the object moved to the second IR obstacle detector sensor, the second set of High LEDs are glowing and the first set again turns to DIM state (Fig. 8d). These results show the efficiency of proposed idea and gives the immediate validation of the proposed model.

In addition, Fig. 9 illustrates the total number of objects / vehicles passed through the road and the derived results of Fig. 8 are summarized in table 2.

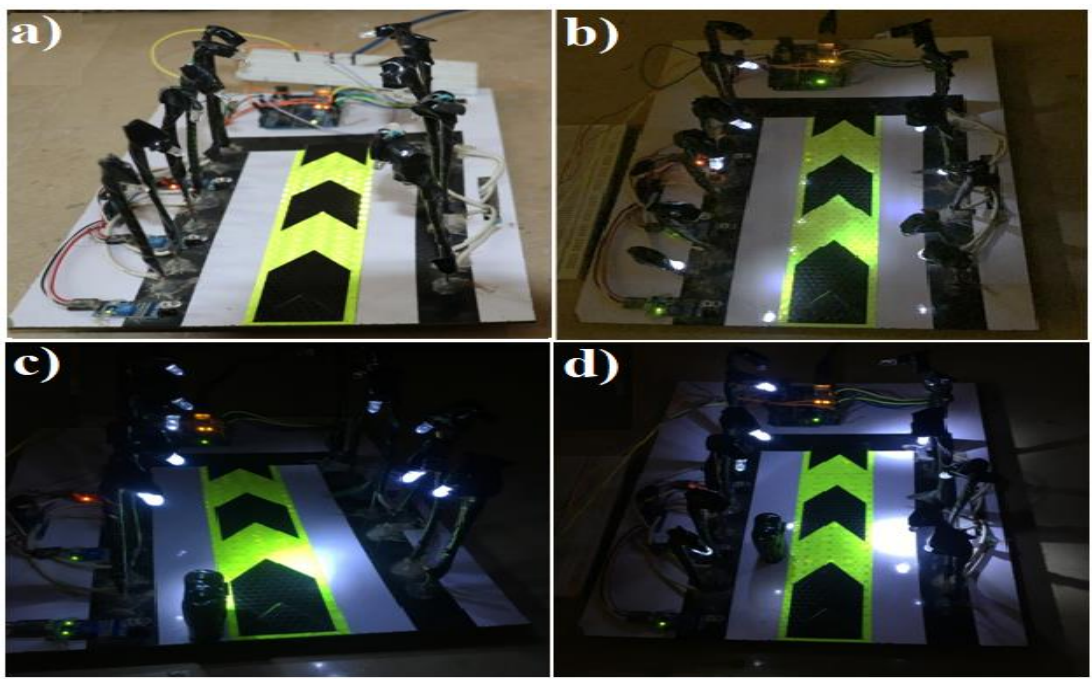

**Figure 8.** Result diagrams of automatic streetlight control system that turn to DIM at night and HIGH on object detection. (a) Shows it is a day-time, so LEDs are not glowing. (b) Shows it is a night-time and Dim LEDs are glowing. (c) Shows object in-front of first IR sensor and first set of High LEDs are glowing while remaining are in DIM mode. (d) Shows motion in-front of second IR sensor so, only second set of LEDs are glowing HIGH and others are in DIM state.

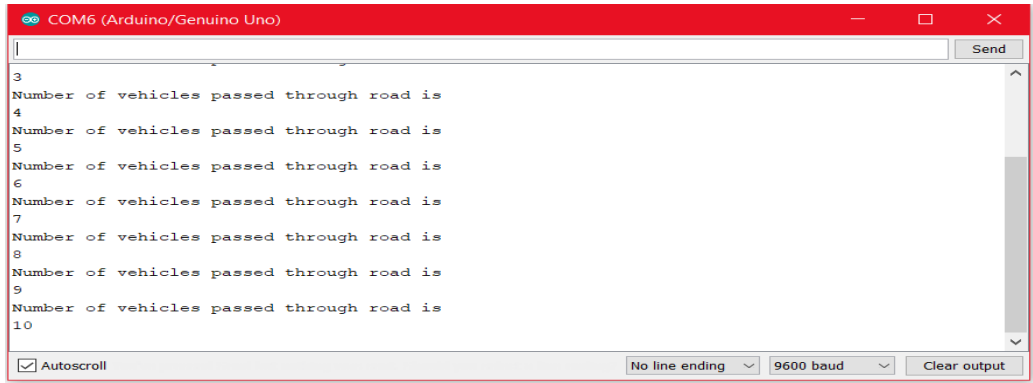

**Figure 9.** Serial monitor output according to traffic flow.

| Device Name | Input Data | Verified Results | Remarks |
|---|---|---|---|
| Arduino Board testing | Digital Signal | Switching of LEDs at different intervals | Hardware is accurate |
| Light Dependent Resistor testing | Outside light intensity values | Dim/High LEDs glows according to light intensity and noted on the Serial monitor | Hardware is accurate |
| IR Obstacle Senor testing | Sense Motion | High LEDs glows whenever it detects motion | Hardware is accurate |

**Table 2.** Derived results after implementation.

**C. Object Dependent Automation System**

As per our motive, the idea of this paper is to create such innovation for our current street light system so that the power consumption can be saved. As presented in Fig. 8, when there are no vehicles on the road at night-time, still the dim light continuously glows, and it wastes energy. So, we enhanced our task with the switching of the street lights based on the IR Obstacle detection sensor. In which when the object is detected at night then the LEDs will switch ON automatically, otherwise the lights will remain OFF. This task is implemented on another board and the circuit design can be seen in Fig. 10. In addition, there is also an automatic door system in this design that will operate with motor and IR obstacle detector sensor. The motor will automatically open the door when IR sensor detects an authorized vehicle in front of the door and shut it when no vehicles are detected.

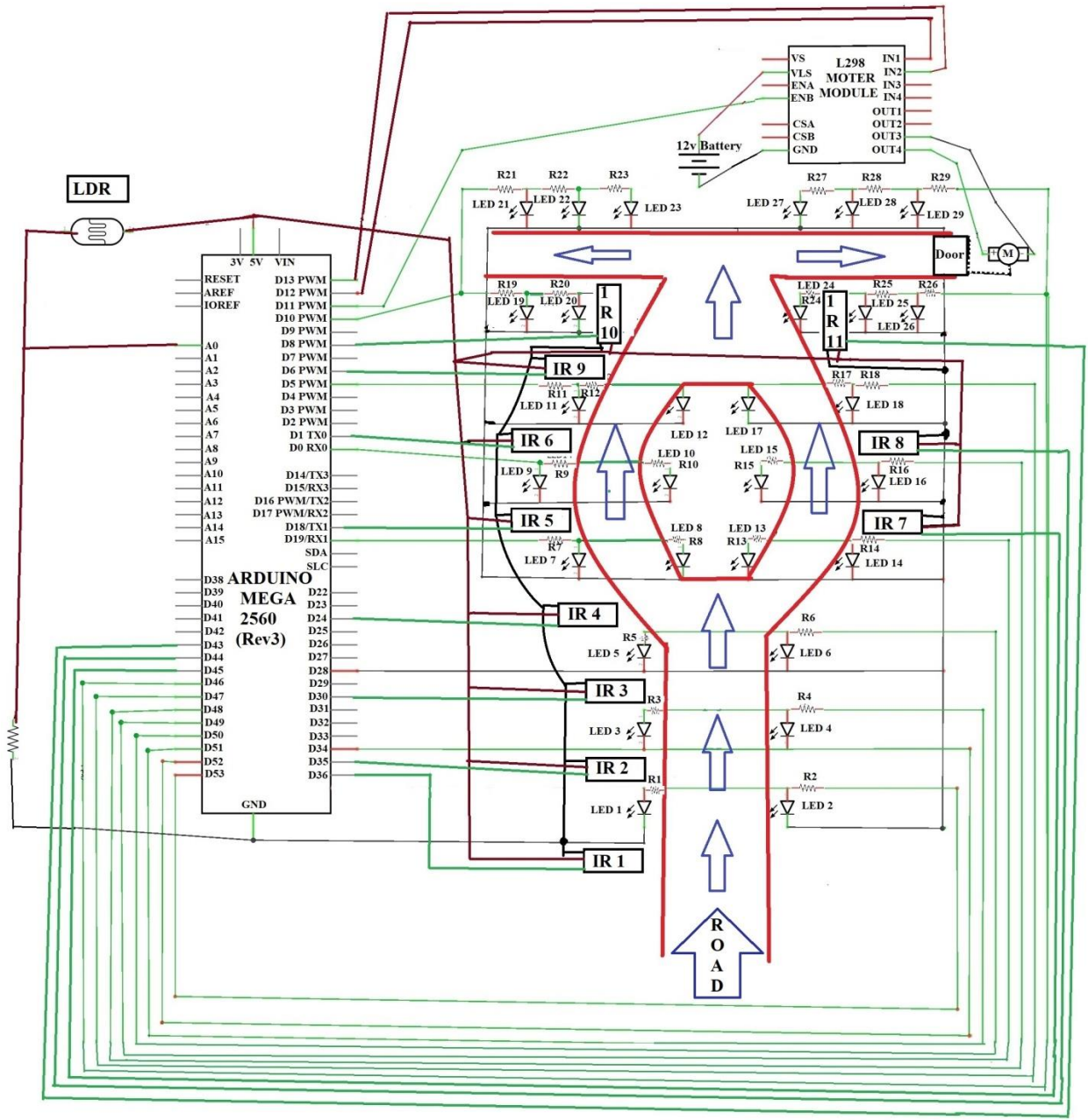

**Figure 10.** Circuit design of enhanced project with no Dim LEDs capability and having automatic door system.

Fig. 10 shows circuit design of automatic streetlight control system without Dim LEDs capability based on Vehicle Detection Using Arduino Mega with an automatic door system. In this, 1 LDR sensor, 29 LEDs, 30 resistors and 11 IR obstacle detector sensors, 1 DC Motor, 1 L298 Motor Module and 1 Arduino Mega have been used. In this method, one leg of LDR sensor is connected to Arduino analog pin number A0 and another leg to VCC pin and same with a resistor to the ground port of Arduino. The threshold value is adjusted to 10 from a discrete value (0-1023) for understanding whether it's DAY or NIGHT. All the positive terminals of the LEDs are connected along with resistors to pin number 0, 5, 10, 19 and 46 - 53as per the circuit diagram given in Fig. 10. In addition, the ground of all the LEDs are connected to Ground port.

The IR obstacle avoidance sensors are connected to the Arduino port from pin number 1, 6, 8, 18, 24, 30, 35, 36 and 43 - 45 respectively, which is the input signal to the Arduino board. The ground of all the IR obstacle avoidance sensors are connected to GND port and all VCC of IR obstacle avoidance sensors are directed in Arduino 5V pin. Initially, set the IR obstacle avoidance sensors to HIGH at the start if there is no motion. Finally, the OUT-pin3 is connected with the L298 motor module with one side of DC motor and OUT-pin4 to another side. In this regard, pin B is enabled and IN-pin 1, 2 is further attached to pin number 11, 12, 13 of Arduino respectively. To supply the motor, 12v pin to the positive terminal and GND to the negative terminal of a 12v battery.

Fig. 11 shows the result diagrams of automatic street lights that turn to ON/OFF only without Dim capability on vehicle movement at night-time and having an automatic door system using Arduino Mega. In this way, Fig. 11a is represented the day-time with no LEDs are glowing. On the other hand, in Fig. 11b, the first set of High LEDs are glowing in the nighttime because there is a motion detected by the first IR sensor. Similarly, when the object moved forward from second IR obstacle detector sensor then the third IR obstacle detector sensor detect motion and third set of High LEDs are glowing, last set and remaining LEDs are turned OFF (Fig. 11c). In this task, when a vehicle is detected in front of the door by IR obstacle detection sensor, it will automatically open and corresponding lights will turn ON, can be seen in Fig. 11d.

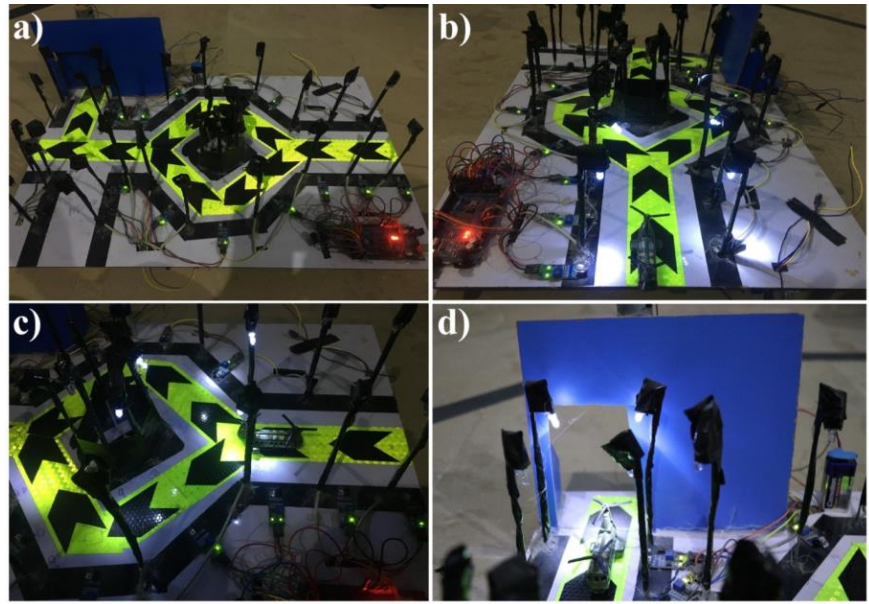

**Figure 11.** Result diagrams of enhanced work with automatic door system and only ON/OFF capability. (a) Shows it is a day-time, so LEDs are not glowing. (b) Object in-front of first IR sensor so, High LEDs are glowing there. (c) Shows motion in-front of third IR sensor so, third set of LEDs are glowing. (d) Shows a vehicle detected in-front of the door, so it is automatically opened and concerned LEDs are glowing.

## III. CONCLUSION

The proposed streetlight automation system is a cost effective and the safest way to reduce power consumption. It helps us to get rid of today's world problems of manual switching and most importantly, primary cost and maintenance can be decreased easily. The LED consumes less energy with cool-white light emission and has a better life than high energy consuming lamps. Moving to the new & renewable energy sources, this system can be upgraded by replacing conventional LED modules with the solar-based LED modules. With these efficient reasons, this presented work has more advantages which can overcome the present limitations. Keep in mind that these long-term benefits; the starting cost would never be a problem because the return time of investment is very less. This system can be easily implemented in street lights, smart cities, home automation, agriculture field monitoring, timely automated lights, parking lights of hospitals, malls, airport, universities and industries etc.